\documentclass[a4paper,fleqn]{cas-sc}

\usepackage[authoryear]{natbib}

\usepackage{siunitx}

\def\tsc#1{\csdef{#1}{\textsc{\lowercase{#1}}\xspace}}
\tsc{WGM}
\tsc{QE}

\newcommand{\degre}{^\circ}

\newcommand{\inc}{I}

\begin{document}
\let\WriteBookmarks\relax
\def\floatpagepagefraction{1}
\def\textpagefraction{.001}

\shorttitle{Dynamical evolution of the Uranian satellite system III}    

\shortauthors{S.R.A. Gomes \& T. Keizer}   

\title [mode = title]{Dynamical evolution of the Uranian satellite system}
\title [mode = sub]{III. The passage through the $7/4$ MMR between Miranda and Ariel}

\author[1]{S\'ergio R.A. Gomes}[orcid=0000-0001-9386-3996]
\cormark[1] 
\ead{s.r.alvesgomes@uva.nl} 

\affiliation[1]{organization={Anton Pannekoek Institute},
            addressline={Science Park 904}, 
            city={Amsterdam},
            postcode={1090 GW}, 
            country={Netherlands}}

\cortext[cor1]{Corresponding author}

\author[1]{Tibi Keizer}

\begin{abstract}
The passage through the $5/3$ mean-motion resonance between Ariel and Umbriel, two of Uranus's largest moons, still raises several open questions. Previous studies suggest that, in order to reproduce the current orbital configuration, Ariel must have had an eccentricity of approximately $\sim 0.01$ before the resonance encounter, which would prevent resonant capture. However, the rapid tidal circularization of Ariel's orbit implies that some prior mechanism must have excited its eccentricity before the resonance encounter.
In this work, we performed a large number of simulations using an $N$-body integrator to assess whether the earlier $7/4$ mean-motion resonance between Miranda and Ariel could serve as a mechanism to increase Ariel's eccentricity. Our results show that, due to divergent migration, resonance capture does not occur. As the satellites cross the nominal resonance, Ariel's eccentricity is only excited to $\num{3.4e-4}$, substantially smaller than the required value.
Therefore, the $7/4$ mean-motion resonance is not a viable mechanism for increasing Ariel's eccentricity.
\end{abstract}


\maketitle

\section{Introduction}\label{sec:intro}

The intricate dynamical evolution of Uranus's five regular satellites has attracted increasing interest in recent years. In particular, the possibility of past passages through mean-motion resonances (MMRs) has been the subject of considerable attention. At present, none of the satellites are locked in resonance. This implies that any previous resonances were either crossed without capture or escaped from after temporary entrapment.  

The most recent low-order MMR thought to have been crossed is the $5/3$ MMR between Ariel and Umbriel. \citet{Tittemore_Wisdom_1988} investigated this resonance passage within the planar approximation and reported that, in order to prevent stable long-term capture, the eccentricities of the satellites should be at least $\sim 0.01$. Later, \citet{Cuk_etal_2020} revisited this problem in three dimensions. Their results indicate that resonance capture not only excites the eccentricities of Ariel and Umbriel, but also significantly increases their inclinations to values far larger than those currently observed. 

More recently, \citet{Gomes_Correia_2023,Gomes_Correia_2024I} applied a secular 3D model to the system, and proposed that in order to reproduce the present orbital architecture of the system, the satellites must avoid long-term capture in the $5/3$ Ariel-Umbriel MMR. To achieve that, Ariel must have maintained an eccentricity of at least $0.015$ before the resonance encounter. Such a large eccentricity would require an earlier excitation mechanism, as tidal dissipation acting on the satellites efficiently damps the orbit toward its equilibrium value of $\num{1.5e-4}$  \citep{Gomes_Correia_2024I}.  

In this note, we investigate one possible mechanism for providing such orbital excitation: the $7/4$ MMR between Miranda and Ariel. We perform a series of numerical simulations with the $N-$body code \texttt{TIDYMESS} \citep{Boekholt_Correia_2023} to study the dynamical consequences of the resonance crossing, focusing on its impact on the orbital evolution of Ariel.  

The paper is structured as follows: in Sect. \ref{sec:theoretical_background} we outline the theoretical framework of this study; in Sect. \ref{sec:numerical_setup} in develop the setup for the numerical simulations; in Sect. \ref{sec:results} we present the results; in Sect. \ref{sec:conclusion} we summarize and discuss the results.
\begin{table}
    \centering
    \caption{Physical and mean orbital parameters of Uranus, Miranda, and Ariel, obtained from \cite{Gomes_Correia_2024I}. }
    \label{tab:orbital_physical_properties_satellites}
\renewcommand{\arraystretch}{1.1}
    \begin{tabular}{c|c|c|c}
		\hline       
                                                    & Uranus              & Miranda         & Ariel           \\
       \hline
       Mass \\ ($\rm \times 10^{-10} \, M_{\odot}$) & $\num{4.365628e5}$  & $0.323997$      & $6.291561$      \\
       Radius (km)                                  &  $25\,559$          & $235.8$         & $578.9$         \\
       $J_2$                                        &  $\num{3.5107e-3}$  & $\num{6.10e-3}$ & $\num{1.39e-3}$ \\
       $k_2$                                        &  $\num{0.104}$      & $\num{8.84e-4}$ & $\num{1.02e-2}$ \\
       $\zeta$                                      & $\num{0.225}$       & $\num{0.327}$   & $\num{0.320}$   \\
       Period (day)                                 &  $0.718328$         & $1.413480$      & $2.520381$      \\
       $a \, (R_0)$                                 &                     & $5.080715$      & $7.470167$      \\
       $e \, (\times 10^{-3})$                      &                     & $1.35$          & $1.22$          \\
       $\inc \, (\degre)$                           &                     & $4.4072$        & $0.0167$        \\
       \hline
    \end{tabular}
    \renewcommand{\arraystretch}{1.0}
    \end{table}

\section{Theoretical Background}\label{sec:theoretical_background}

The current orbital evolution of Uranus's major satellites is primarily governed by tidal interactions between the satellites and the planet. 
In the widely used weak friction tidal model \citep[see][]{Gomes_Correia_2023,Gomes_Correia_2024II}, a body deforms into a tidal equilibrium shape with a small time lag ($\tau$) in response to the gravitational pull of another object due to friction within the perturbed body. This framework is particularly suitable for systems with low eccentricities and moderate tidal dissipation, where the assumptions of a small tidal lag and linear response remain valid. 

This dissipative effect leads to damping of both the orbital eccentricity ($e$) and inclination ($\inc$) of the satellites, as well as orbital migration. This is given by \citep{Gomes_Correia_2024I}
\begin{equation}\label{eq:semi_major_axis_migration}
   \frac{\dot{a}_k}{a_k} = k_{2,0} \tau_0 
   \frac{6 \mathcal{G} m_k^2 R_0^5}{\beta_k a_k^8}  
   \left(\frac{\omega_0}{n_k} - 1\right) \ ,
\end{equation}
where the subscript $0$ refers to the planet, and the subscripts $k=1,2...$ refers to the satellites in ascending order of distance to the planet. $\mathcal{G}$ is the gravitational constant, $a_k$ is the semi-major axis, $m_k$ is the mass, $n_k$ is the mean motion, $\beta_k = m_0 \ m _k / (m_0 + m_k ) $ the reduced mass, $R_0$ the planets' radius, $k_{2,0}$ the planets' Love number, $\tau_0$ the planets' tidal time lag and $\omega_0$ the planetary spin rate.  

Due to their distinct values of $a_k$ and bulk properties, each satellite migrates at a different rate. Consequently, their mutual separations evolve over time. This evolution may bring them into particular mean-motion resonance configurations of the form  
\[
   (p+q) \, n_2 + q \, n_1 \approx 0 \ ,
\]  
where $n_1$ and $n_2$ are the mean motions of the inner and outer satellites, respectively, and $p$ and $q$ are integers, where $q$ is the order of the resonance. For the case of the $7/4$ MMR, $p = 4$ and $q = 3$.

While in or near the resonance, mutual gravitational perturbations no longer cancel over time, and both the eccentricities and inclinations can be significantly excited. The migration rates of the satellites also change in such a way that their orbital period ratios remain locked in the nominal resonant value. The resonance can then be broken due to resonance overlapping, where multiple resonance solutions overlap and lead to chaotic evolution. Or, if the resonance is approached with sufficiently large eccentricity or inclination, resonance entrapment is prevented and the system only fells a small perturbation.
 
However, resonance capture in two-body MMR require convergent migration. That is, that the ratio between the mean-motions of the two bodies must be diminishing. Nevertheless, even without resonance capture, the near resonance motion is still sufficient to induce some excitation on the orbits of the bodies.

For a more detailed theoretical review, we refer the reader to \cite{Gomes_Correia_2023,Gomes_Correia_2024II}.


\section{Numerical setup}\label{sec:numerical_setup}

Due to the chaotic and irreversible nature of resonant interactions, backward integration through an MMR is inherently unfeasible. Therefore, we need to initialised the system just before nominal resonant value $7/4$, and then integrate forward in time. 

Following the same methodology as \cite{Gomes_Correia_2023,Gomes_Correia_2024II}, we calculated the pre-MMR $a_k$ using Eq.~\ref{eq:semi_major_axis_migration}. From the current orbits of the satellites and $\omega_0$ (Table~\ref{tab:orbital_physical_properties_satellites}), we integrated $a_k$ backward up to the nominal $7/4$ values.   For consistency we adopted the same 
$k_{2,0} \tau_0 = 0.064 $ s, 
$k_{2,1} \tau_1 = 0.034 $ s, and
$k_{2,2} \tau_2 = 0.707 $ s, 
as \cite{Gomes_Correia_2024II}, that is, a tidal quality factor of $Q_0 = 8\,000$.  In addition, to guarantee that the simulations start before the resonance, we slightly increased Ariel's semi-major axis such that 
\begin{equation}
  \begin{split}
    a_{Miranda} = & 5.035\,927  R_0 \quad \text{and}\\
    a_{Ariel} = & 7.312\,222 R_0 \, .   
\end{split}
\end{equation}

We conducted $2\,000$ simulations with initial eccentricities set to zero. Due to the mutual perturbations between Miranda and Ariel, the eccentricities are expected to grow to the equilibrium value. 
Since we are following the assumption of \cite{Gomes_Correia_2024II} that the 5/3 Ariel-Umbriel MMR must be skipped, we do not expect the inclinations to significantly change over time. Thus, the initial inclinations were set to current values (Table \ref{tab:orbital_physical_properties_satellites}).

To ensure an uniform distribution of the resonance angle $7 \lambda_2 - 4 \lambda_1$, where $\lambda$ is the mean longitude, the mean anomalies, the longitudes of the pericenter and  the longitudes of the ascending node of both satellites where randomly sampled between $0$ and $2\pi$ radians.
Finally, the obliquities of all bodies where set to zero, and the second order gravitational fields $J_2$ and normalized moments of inertia $\zeta$ are given in Table \ref{tab:orbital_physical_properties_satellites}.

\cite{Gomes_Correia_2024I} conclude that the general evolution of the orbital elements is not sensitive to the enhancement of the tidal strength. As a result, the simulations can be sped up by a factor of at least $\times 1\,000$ without compromising the accuracy of the results. We also conducted a convergence test and reached the same conclusion.

Then, each simulation was integrated over 20 Myr, using \texttt{TIDYMESS} \citep{Boekholt_Correia_2023}, which enables detailed simulations of the system's dynamical evolution. It self-consistently accounts for gravitational interactions between bodies, spin dynamics, tidal deformation, and shape evolution, while also incorporating complex physical effects such as rotational flattening. It approximates body shapes as ellipsoids and includes gravitational potentials up to quadrupole order.

\section{Results}\label{sec:results}

The first result is that the forced equilibrium values for the satellites were also very low, around $\num{5.5e-4}$ and $\num{1.5e-4}$, respectively. This values are consistent with the ones reported in \cite{Gomes_Correia_2024I}.

By computing the tidal migration ratios of both satellites, it is clear that the migration of Miranda and Ariel is divergent. Thus, resonance capture is not expected (Sect. \ref{sec:theoretical_background}). In fact, we did not observe resonance capture in none of the simulations performed.
However, in the vicinity of the nominal value of the $7/4$ ratio, we observe a small perturbation on eccentricity values of both satellites.

\begin{figure}
  \centering
  \includegraphics[width = 0.8\linewidth]{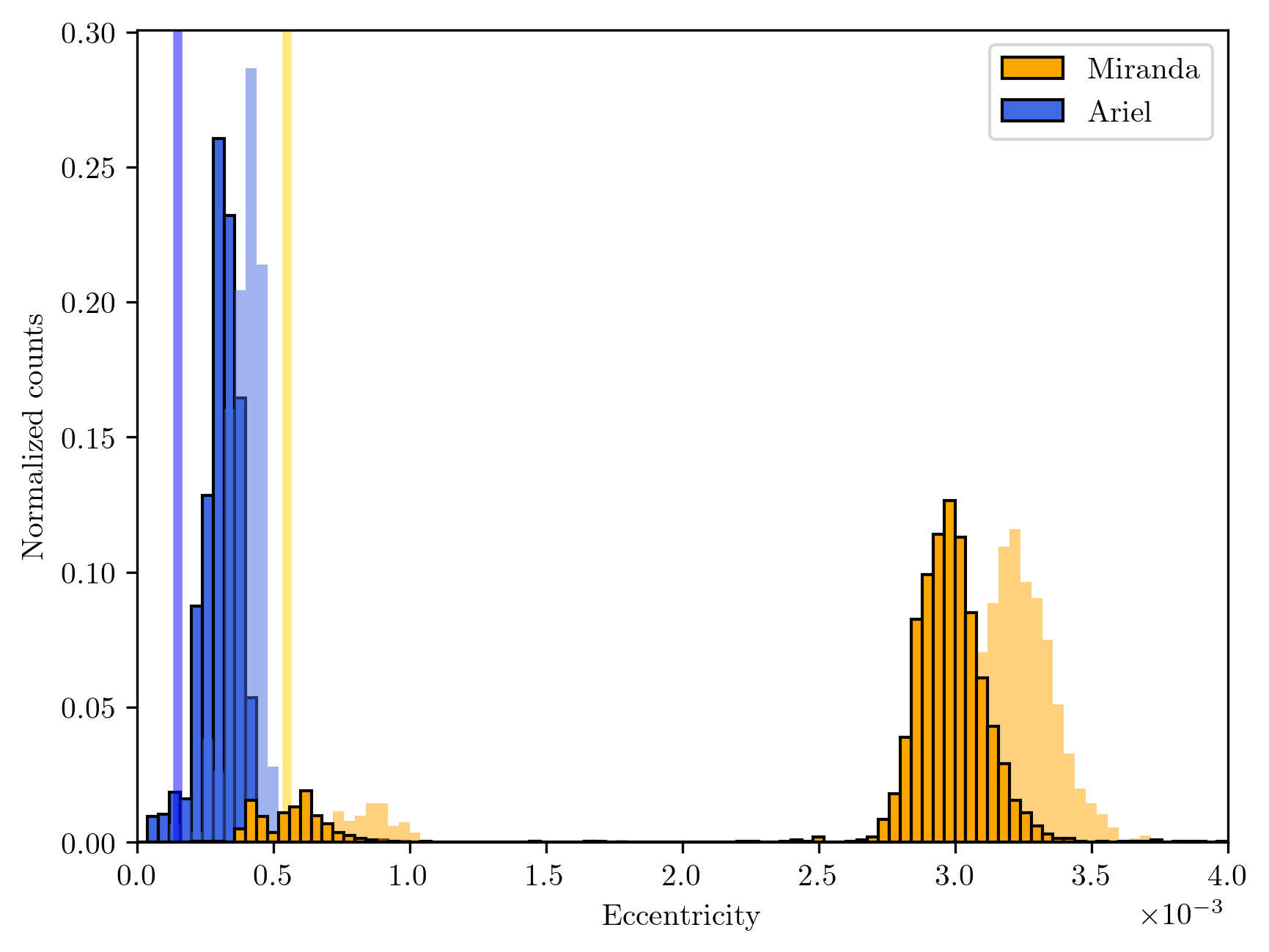}
  \caption{Distribution of the mean eccentricities (dark shaded bins) and maximum free eccentricities (light shaded bins) for Miranda (orange) and Ariel (blue), after the passage through the 7/4 MMR.
  The thin yellow and blue lines mark the mean equilibrium eccentricities values before the resonance passage for Miranda and Ariel, respectively.}
  \label{fig:histogram}
\end{figure}

In Fig.~\ref{fig:histogram} we present the mean eccentricity values after the nominal resonance crossing for both Miranda (orange) and Ariel (blue). In lighter colours, we also present the corresponding maximum free eccentricity values. The mean eccentricity values before the resonance encounter are marked as vertical thin lines with matching colours. For a better comparasion, we also fitted a gaussian to the distribution of eccentricities of Ariel, and to the right hand pile of the distribution of Miranda (see Table \ref{tab:gaussian_distributions}).

For Ariel, we observed that the final values cluster around a mean of $\num{3.1e-4}$ with a standard deviation of $\num{7.0e-5}$. This indicates that resonance passage induces an increase in the mean eccentricity by approximately a factor of 8. Regarding the maximum free eccentricity, the distribution peaks at a slightly larger value; however, these values remain two orders of magnitude lower than the target value required to bypass the 5:3 Ariel–Umbriel mean-motion resonance (MMR).

For Miranda, due to its smaller size, the initial equilibrium eccentricity is larger than that of Ariel. The resonance crossing induces much larger perturbations in the satellite’s eccentricity. In contrast to Ariel, we observe a broader, bimodal distribution, with some orbits clustering around the equilibrium value and a significant fraction reaching much higher eccentricities, with a mean of $\num{3.0e-3}$ and a standard deviation of $\num{1.2e-4}$..

We also analysed the final distribution of the inclinations and did not observe any measurable excitation of those values.

In order probe the impact of the aceleration factor used to speed up the simulations, we also conduted a 100 simulations with an aceleration factor of $\times 100$ and $\times 10 $, each.  In Table \ref{tab:gaussian_distributions} we present the mean values and standard deviation of the gaussian fit to the distribution of final mean eccentricities for both satellites. The mean values are almost identical for the three aceleration factor, with just an increase of the standar deviation, most likelly due to the smaller number of simulations.

Finally, we conducted 300 simulations with an aceleration factor of $\times 1000$, but including Umbriel, Titania, and Oberon. We extrapolated the semi-major axis of the satellites using the same method described in Sec. \ref{sec:numerical_setup}, and the physical and tidal properties were taken from Table 1, from \cite{Gomes_Correia_2024I}. The mean value of the gaussian distribution (Table \ref{tab:gaussian_distributions}) is slightly larger, as well as wider. However, is still in the well within the same order of magnitude and consistent with the values obtained with only two satellites. For Miranda, the distribution is to wide for a gaussian fit, with the values also ranging up to $\num{3.5e-3}$.


\begin{table}
  \begin{center}
  \caption{Gaussian fits to the distributions of final eccentricities obtained from simulations with: two satellites and the central body, using different acceleration factors ($\times 1000$, $\times 100$, and $\times 10$); and five satellites and the central body, using an acceleration factor of $\times 1000$. $\mu$ is the mean value and $\sigma$ the standard deviation of the gaussian.}
  \begin{tabular}{ p{0.7cm} | c c | c c} 
  \hline
   & $\mu_M$ & $\sigma_M$ & $\mu_A$ & $\sigma_A$ \\
  \hline
  $\times 1000$ & \num{3.0e-3} & \num{1.2e-4} & \num{3.1e-4} & \num{7.0e-5} \\
  $\times 100$  & \num{3.0e-3} & \num{1.1e-4} & \num{2.9e-4} & \num{8.5e-5} \\
  $\times 10$   & \num{3.0e-3} & \num{1.1e-4} & \num{3.0e-4} & \num{8.8e-5} \\
  5 sat & -- &  -- & \num{5.5e-4} & \num{2.4e-4} \\
  \hline
  \end{tabular}
  
  \label{tab:gaussian_distributions}
\end{center}
\end{table}

\section{Conclusion}\label{sec:conclusion}

In this work, we aim to assess whether the $7/4$ MMR between Miranda and Ariel can significantly increase Ariel's eccentricity, thereby preventing capture into the subsequent $5/3$ MMR between Ariel and Umbriel. We conducted a series of two-satellite numerical simulations, starting from zero initial eccentricities.

Although resonant capture is not expected due to the divergent nature of the $7/4$ MMR, the passage through this resonance can still induce orbital excitation. Adopting ${Q_0 = 8\,000}$ and ${k_{2,0} = 0.103}$, the $7/4$ MMR occurs approximately $620$ Myr before the $5/3$ Ariel-Umbriel MMR (Eq.~\ref{eq:semi_major_axis_migration}). The eccentricity damping timescale for Ariel is $\sim 260$ Myr \citep{Gomes_Correia_2024I}. This implies that, in order to have a measurable likelihood of skipping the $5/3$ MMR, Ariel's eccentricity after leaving the $7/4$ MMR must reach $\sim 0.06$. However, our results demonstrate that the eccentricity excitation induced by the passage through the $7/4$ MMR is insufficient to raise Ariel's eccentricity to the required value. The resonance crossing induces only a small perturbation, increasing Ariel's eccentricity to $\num{3.1e-4}$, which is expected to be rapidly damped back to its equilibrium value.
Therefore, the 7/4 MMR between Miranda and Ariel cannot account for the $\sim 0.01$ eccentricity required for Ariel to avoid entrapment in the $5/3$ MMR with Umbriel. 

We also conducted simulations using smaller acceleration factors and including all five of the largest Uranian satellites, and found no — or only negligible — differences, thereby reinforcing the robustness of our results.

This suggests that an alternative dynamical mechanism must be responsible, possibly involving a different dynamical event, such as the ${3 n_1 - 8 n_2 + 4 n_3}$ three-body MMR between Miranda, Ariel, and Oberon, or the $4 n_1 - 10 n_2 + 5 n_3$ three-body MMR between Miranda, Ariel, and Umbriel, both occurring $250$–$300$ Myr before the $5/3$ MMR.

Another possibility is that the Ariel experience resonance locking with Uranus. Such would  decrease the quality factor of planet to bellow $Q_0 = 1\,000$ \citep{Nimmo_2023,Jacobson_Park_2025}, substantially increasing the migration rate of Ariel. This could lead to a significantly different orbital evolution, making scenarios such as the $2/1$ MMR between Ariel and Umbriel—previously considered very unlikely to escape—now dynamically feasible \citep{Rossi_etal_2025}.

\section*{Data availability}
The data necessary to reproduce the figures in this paper are publicly available on Zenodo:
\href{https://doi.org/10.5281/zenodo.17348764}{10.5281/zenodo.17348764}.

\section*{Acknowledgements}
This work was support by NWO, through the Stimuleringbeurs. 
We thank the two anonymous reviewers for their valuable comments, which helped improve this manuscript. We also acknowledge Antonija Oklopčić, Silvia Tonnen, Giacomo Lari, Mattia Rossi, and Mélanie Sailenfest for their insightful discussions and support.

\bibliographystyle{cas-model2-names}

\bibliography{bibliography}

\end{document}